# Fast, High Resolution and Wide Modulus Range Nanomechanical Mapping with Bimodal Tapping Mode

*Marta Kocun, Aleksander Labuda, Waiman Meinhold, Irène Revenko and Roger Proksch**

Asylum Research - an Oxford Instruments Company, Santa Barbara, CA  93117

**ABSTRACT**

Tapping mode atomic force microscopy (AFM), also known as amplitude modulated (AM) or AC mode, is a proven, reliable and gentle imaging mode with widespread applications. Over the several decades that tapping mode has been in use, quantification of tip-sample mechanical properties such as stiffness has remained elusive. Bimodal tapping mode keeps the advantages of single-frequency tapping mode while extending the technique by driving and measuring an additional resonant mode of the cantilever. The simultaneously measured observables of this additional resonance provide the additional information necessary to extract quantitative nanomechanical information about the tip-sample mechanics. Specifically, driving the higher cantilever resonance in a frequency modulated (FM) mode allows direct measurement of the tip-sample interaction stiffness and, with appropriate modeling, the setpoint-independent local elastic modulus. Here we discuss the advantages of bimodal tapping, coined AM-FM imaging, for modulus mapping. Results are presented for samples over a wide modulus range, from a compliant gel (~100 MPa) to stiff materials (~100 GPa), with the same type of cantilever. We also show high-resolution (sub-nanometer) stiffness mapping of individual molecules in semi-crystalline polymers and of DNA in fluid. Combined with the ability to remain quantitative even at line scan rates of nearly 40 Hz, the results demonstrate the versatility of AM-FM imaging for nanomechanical characterization in a wide range of applications.

**Corresponding Author**

*Address correspondence to roger.proksch@oxinst.com

The function and performance of many emerging materials and devices depend on their nanoscale morphology, composition and mechanical properties. However, as the length scales of these materials and devices decrease, quantitative measurements of parameters such as the elastic modulus become increasingly challenging. Thus, a nanomechanical mapping technique—an imaging technique that provides high spatial resolution simultaneously with quantitative mechanical properties—has been a long-standing goal for many fields. Atomic force microscopy (AFM) is well suited for nanoscale characterization of surfaces and interfaces, with image sizes that can range from tens of micrometers down to fractions of a nanometer. Additionally, the nondestructive nature of AFM, together with its minimal sample preparation requirements, makes it exceptionally versatile.

One of the most common measurement modes in AFM is the dynamic mode called amplitude modulation (AM-AFM), also known as tapping mode or AC mode. During AM-AFM imaging, the cantilever is excited dynamically near a resonant frequency, and the amplitude and phase of the cantilever oscillation are measured. The amplitude is used as a feedback signal for controlling the tip-sample distance in order to track sample topography. Advantages of AM-AFM include minimal tip and sample damage (which allows imaging of softer samples compared to contact mode and force mapping modes), low lateral forces, and high-resolution and very fast scan imaging capabilities.[1,2] A significant drawback of AM-AFM, which drives the cantilever at a single frequency, is its difficulty in quantifying sample mechanical properties. AM-AFM can only provide the ratio of storage to loss modulus, because the phase response is affected by both conservative and dissipative interactions.[3–6]

Roughly a decade ago, it was recognized that simultaneously exciting higher resonance(s) of the cantilever provides additional information regarding the tip-sample interactions.[7–11] One variation of this approach is called bimodal imaging, where two resonant modes are excited simultaneously. Bimodal AFM can be performed in different modes of operation, such as amplitude modulation (AM),[12–16] phase modulation (PM)[17–20] and frequency modulation (FM).[21–25] Originally bimodal experiments were performed in the AM-AM configuration,[8] while the FM-FM configuration was later explored by Heruzzo *et al.*[2]

Here, we demonstrate the capabilities of a hybrid bimodal technique for nanomechanical mapping that combines the robust simplicity of AM operation for the first resonance of the cantilever with the high sensitivity and signal-to-noise ratio of FM operation for the second resonance. This "AM-FM" imaging mode allows for a large dynamic range of measurable modulus, molecular-level spatial resolution at very fast scan speeds and quantitative mapping of moduli, stiffness and indentation depths.

**RESULTS AND DISCUSSION**

**Bimodal AM-FM Imaging**

The principles of AM-FM operation are described in Figure 1. Prior to tip-sample interaction, the first cantilever eigenmode is excited near its resonant frequency $f_1$ with a large amplitude (typically, $A_{1,\text{free}} \approx 100$ nm), and the second (or higher) eigenmode is excited near its resonant frequency $f_2$ with a much smaller amplitude (typically, $A_{2,\text{free}} \approx 1$ nm). In the experiments described here, the driving force was provided by modulating the power of a blue laser focused on the base of the cantilever. The spring constants $k_1$ and $k_2$ and quality factors $Q_1$ and $Q_2$ of the first and second eigenmodes, respectively, are determined before the experiment by fitting the thermal response of the cantilever away from the surface.

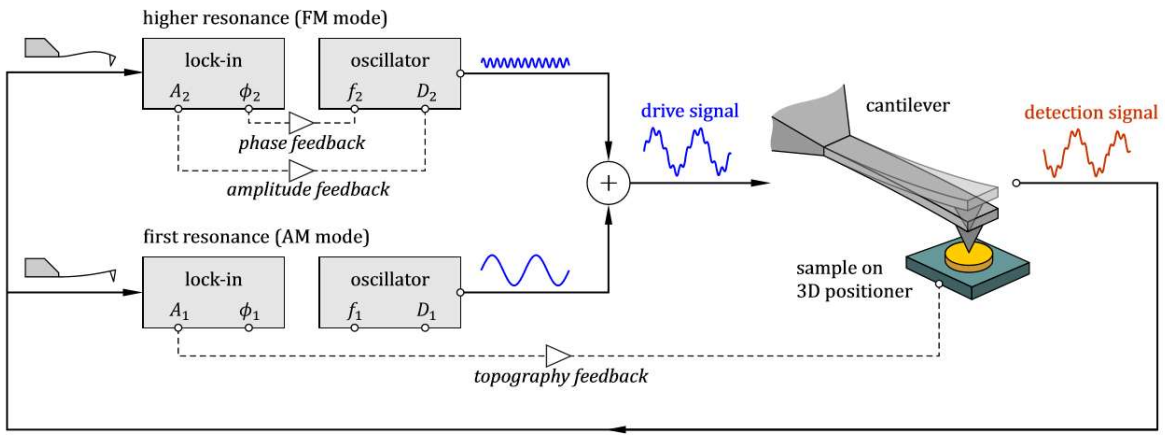

**Figure 1.** System diagram for bimodal AM-FM imaging mode. The first resonance of the cantilever is driven with a large amplitude $A_1$, and a higher resonance is driven simultaneously with a small amplitude $A_2$. The topography feedback loop uses the $A_1$ to track topography. The amplitude feedback loop maintains constant oscillation amplitude of the higher resonance. The phase feedback loop tracks the frequency of the higher resonance to drive it on resonance despite changes in sample modulus.

Upon approaching the sample, the deflection signal of the interacting cantilever is analyzed by a lock-in amplifier to determine the amplitude $A_1$ and phase $\phi_1$ response of the first eigenmode. The topography feedback adjusts the sample height to maintain a constant, predetermined amplitude setpoint $A_{1,\text{set}}$. This type of feedback is identical to that used in AM-AFM imaging for topography tracking.

The deflection signal is also analyzed by a second lock-in at the frequency of the higher eigenmode. Here, a frequency feedback loop continuously adjusts the excitation frequency by a small amount $\Delta f_2$ to maintain the second eigenmode on resonance as the tip interacts with the sample. In addition, an amplitude feedback loop is used to maintain constant oscillation amplitude of the second eigenmode. This ensures that the small amplitude of the second mode remains above the detection noise floor while remaining small relative to the indentation depth, which is required by bimodal AFM theory.[26,27]

The AM-FM signals produced by this imaging feedback scheme can be used[26] to extract the maximum indentation depth $\delta$

$$\delta = \frac{3}{4}\frac{k_1}{Q_1}A_{1,\text{free}}\cos\phi_1 \left(\frac{f_2}{2k_2\Delta f_2}\right) \quad (1)$$

and the effective storage modulus of the interaction $E_{\text{eff}}$

$$E_{\text{eff}} = \sqrt{\frac{2}{3}}\pi/R\left(\frac{k_1}{Q_1}\frac{A_{1,\text{free}}}{A_{1,\text{set}}}\cos\phi_1\right)^{-1/2}\left(\frac{2k_2\Delta f_2}{f_2}\right)^{3/2}. \quad (2)$$

These equations assume the AFM tip can be described by a flat punch with radius $R$, which is used for all data presented in this work. Other geometries, such as sphere or cone, are derived elsewhere.[26] Other groups have preferentially used the sphere model for analyzing similar data.[27] In principle, the ability to choose a characteristic tip shape opens the door to an absolute modulus measurement – meaning that armed with well characterized tip shapes and dimensions these models have the potential to allow modulus measurements with no free parameters. In practice, AFM tip shapes are not commonly well characterized so we have taken a different approach of using a material with a "reference" modulus to characterize the tip size parameter (the tip radius $R$). This calibration procedure is explained below.

The key assumption in bimodal AFM theory is that the indentation depth $\delta$ is much smaller than the amplitude $A_1$. This is in fact achieved during AM-FM imaging in practice, where $\delta \approx 0.1\text{-}10$ nm, while typically $A_1 \approx 100$ nm. These are very small indentations depths relative to other mechanical testing techniques. The advantage of such small indentation depths is that the surface can be imaged with very high resolution and with minimal sample damage. The disadvantage of such low indentation depths is that the modulus measurement is more prone to errors and variations caused by surface effects such as roughness, adhesion, and contaminations.

Lastly, the tip-sample interaction stiffness $\langle k_{ts} \rangle$ presented throughout this work is calculated by

$$\langle k_{ts} \rangle = 2RE_{\text{eff}} = \sqrt{\frac{2}{3}} 2\pi \left( \frac{k_1}{Q_1} \frac{A_{1,\text{free}}}{A_{1,\text{set}}} \cos \phi_1 \right)^{-1/2} \left( \frac{2k_2 \Delta f_2}{f_2} \right)^{3/2} \quad (3)$$

where $E_{\text{eff}}$ is calculated as described above for equations (1) and (2).

**Calibration**

Calibration of the cantilever parameters $k_1, k_2, f_1, f_2, Q_1, A_1$ and the tip parameter $R$ in Equations 1 and 2 is required for quantitative analysis of AM-FM data. A procedure for calibrating the stiffness of the cantilever's first resonant eigenmode as well as higher eigenmodes has been outlined in detail in recent work.[28] The approach involves measurements of the cantilever thermal noise and use of the equipartition theorem. This procedure also allows calibration of the cantilever deflection sensitivity (units nm/V) of both resonant eigenmodes, from which the cantilever oscillation amplitudes in nanometers can be determined. The thermal noise measurement also provides a measurement of the frequency and quality factor of both resonances.

An important parameter to determine accurately is the tip size, which here is defined by the punch model tip radius $R$. The simplest calibration approach is to perform an AM-FM measurement on a surface of known modulus prior to the experiment and extract the tip radius value that satisfies eq 2. Any difference between the modulus of this calibration or reference material and that of the sample leads to an increase in dependence on the choice of the assumed contact model (punch, sphere, or cone) used to analyze the AM-FM data. Therefore, it is advisable to minimize the difference in modulus between the reference material and the sample to reduce this model dependence.

As mentioned above, AFM cantilever tips often have ill-defined shapes and sizes. One common method that avoids explicitly characterizing this shape and size is the internal reference method. In this method, an area of the sample with known modulus (typically assumed as the accepted bulk value) is used to calibrate the tip radius. The resulting value of $R$ is then applied to other areas of the sample or even different samples to obtain a modulus map. When using this internal reference method, errors in the other calibration parameters discussed above can be compensated by choosing a tip size scale that imposes the condition the measured modulus matches the reference modulus.

**Quantitative Modulus Measurements**

The ability of bimodal AM-FM imaging to accurately quantify nanoscale modulus can be demonstrated in the context of component identification in a multilayered packaging film. The film was known *a priori* to be composed of four polymers: polyethylene (PE), polypropylene (PP), polyethylene terephthalate (PET) and ethylene vinyl alcohol (EVOH). Calibration of both cantilever resonance spring constants was performed using a procedure outlined elsewhere[28] while the calibration of the tip size was performed on a polystyrene reference sample, where the modulus was assumed to be 3.0 GPa.

Figure 2a shows the modulus map obtained by AM-FM imaging for this sample. By comparing the image to literature values of elastic moduli, the polymer constituents in each region were unambiguously identified by their relative values. The distinction between polymers is clear despite the small relative differences in modulus between the different film components and the indentation depths (<2 nm) shown in Figure 2b. For example, the PP and PET moduli differ by only 20%, yet they can be clearly distinguished in the AM-FM modulus map. The modulus distribution of each component, shown in the histogram in Figure 2c, is consistent with variations in the contact area that are likely linked to the sample roughness.

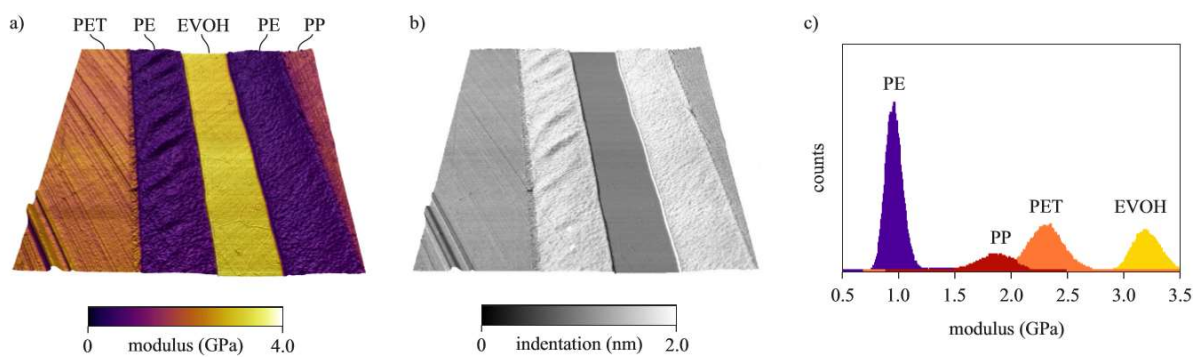

**Figure 2.** Maps of (a) modulus and (b) indentation depth overlaid on topography for a multilayer polymer film. (c) Histogram of the modulus map differentiates the four polymers (PE, PP, PET, EVOH) composing the multilayer film. Calibration of the tip size was performed on a polystyrene reference sample. Cantilever: Olympus AC160 with $k_1 \approx 34$ N/m, $f_1 \approx 293$ kHz, $k_2 \approx 597$ N/m and $f_2 \approx 1.63$ MHz (see Materials and Methods section for definition of variables). Scan size 25 μm.

**Wide Dynamic Range of Modulus Mapping**

In addition to high sensitivity resolution of different materials with similar moduli as shown above, AM-FM imaging can be used to characterize materials across an extremely wide range of modulus. This is demonstrated in Figures 3a-e, which contain AM-FM modulus maps for samples spanning three orders of magnitude in modulus (~0.1 GPa to ~200 GPa). The samples included polydimethylsiloxane (PDMS), thin films of polystyrene/polycaprolactone (PS/PCL) and polystyrene/polyproplyene (PS/PP) polymer blends, lead/tin (Sn/Pb) solder alloy and a patterned titanium (Ti) thin film on silicon (Si). (See Materials and Methods section for details on sample preparation.) Figure 3f summarizes these modulus maps in a single, logarithmic-scale histogram and provides information about the distribution in modulus for sample components.

Remarkably, all the modulus maps in Figures 3a-e were acquired using the same type of cantilever, shown in the inset in Figure 3f. In contrast, most nanomechanical testing techniques require the sensor's spring constant to be roughly matched to the tip-sample stiffness. The high sensitivity of AM-FM can be explained by the fact that changes in the effective cantilever stiffness due to tip-sample interactions are encoded as a frequency shift, which can be very reliably and accurately measured with a lockin amplifier.

A notable example of both the sensitivity and range of this technique is shown in Figure 3d where a 50:50 Sn/Pb alloy solder shows a pattern of softer and stiffer regions in the modulus channel that is not visible in topography (topography not shown; see Ref. 6). Often observed in optical and scanning electron micrographs,[29] this pattern is typical of Sn/Pb solder and is associated with softer, Pb-rich and stiffer, Sn-rich regions of solder. These regions were not identifiable with conventional AM imaging mode with a single resonance: they are not correlated with topographical or phase features; in addition, loss tangent imaging was unable to distinguish them within measurement error.[6] Since the loss tangent is the ratio of the loss and storage moduli $E''/E'$, this lack of contrast may simply be due to these ratios being similar in the two materials or more likely, that the other artifacts as discussed in reference 6 overwhelm any differences in the material ratios. Since AM-FM imaging is sensitive to the storage modulus alone, it is not subject to this limitation and differences in the moduli are clearly visualized, as discussed in reference 6 (see specifically Figure 3 in reference 6 and associated discussion).

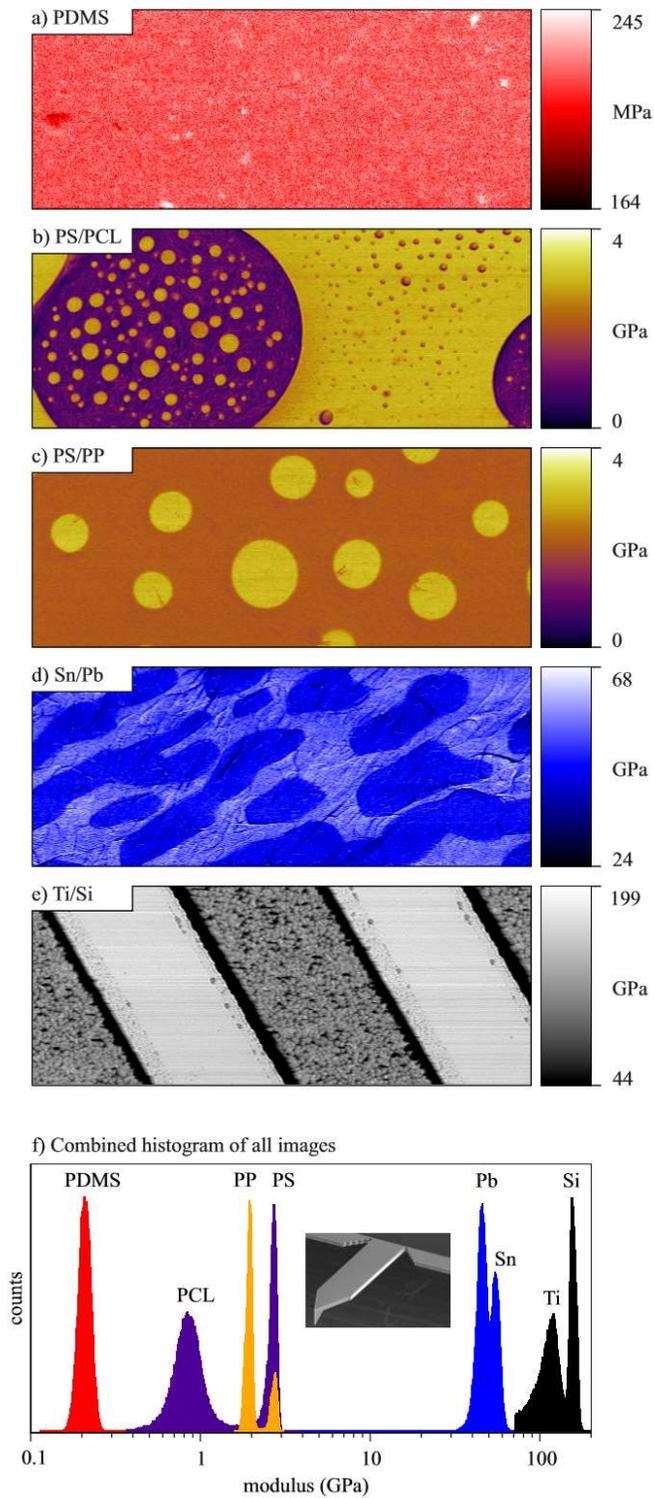

**Figure 3.** AM-FM modulus maps on samples with a wide range of modulus (in order of increasing modulus): (a) PDMS, (b) PS/PCL polymer blend, (c) PS/PP polymer blend, (d) Sn/Pb

solder alloy and (e) Ti thin-film stripes on Si. All images are ~8 μm wide. (f) Combined color-coded histogram of the images in (a)-(e) on a logarithmic scale. The inset shows a scanning electron micrograph of an Olympus AC160 cantilever such as used in all these measurements (160 μm long, nominal $k_1 \approx 30$ N/m).

**Measurements on High-Modulus Materials**

Sub-resonance force measurement techniques such as force-distance curves are often used for modulus measurements with AFM. However, they cannot provide reliable modulus information on very stiff materials due to the relatively low stiffness of commercial AFM cantilevers and the associated shallow indentation depths. In fact, very few AFM modes can quantitatively characterize nanomechanical properties of materials with moduli greater than a few tens of gigapascals.

Contact resonance AFM techniques[30–32] provide the ability to characterize mechanical properties of very stiff samples. However, the downside to such contact mode techniques is that the high normal forces (typically >100 nN) required for imaging can be destructive. This is especially problematic for very high-aspect-ratio features common to the semiconductor industry.

In contrast, the tapping-mode nature of AM-FM bimodal imaging means the tip contacts the sample intermittently, allowing nondestructive imaging and the ability to characterize high modulus samples by virtue of the high stiffness of higher cantilever eigenmodes. An example is shown in Figure 4, where AM-FM modulus mapping was performed on a Si wafer with a regular pattern of ~200-nm-high diagonal Ti. Here, the Si surface was used as an internal reference with an assumed modulus of 165 GPa, which led to a measured Ti modulus of 115 GPa. This measurement is consistent with the bulk modulus of 110-125 GPa.[33]

In addition, Figure 4 demonstrates the independence of the measured modulus on the AM-FM amplitude setpoint as it was varied stepwise during image acquisition. This variation was performed to test the applicability of modulus calculations to the experimental results. Whereas these deliberate changes led to horizontal bands visible in the amplitude and phase images (Figures 4a and b), the calculated modulus map (Figure 4c) does not show any corresponding variations. This implies that both the chosen tip contact model and bimodal theory used to perform the calculation were accurate for this dataset. In this case, the Hertz model with punch tip geometry and radius $R = 3.3$ nm was used.

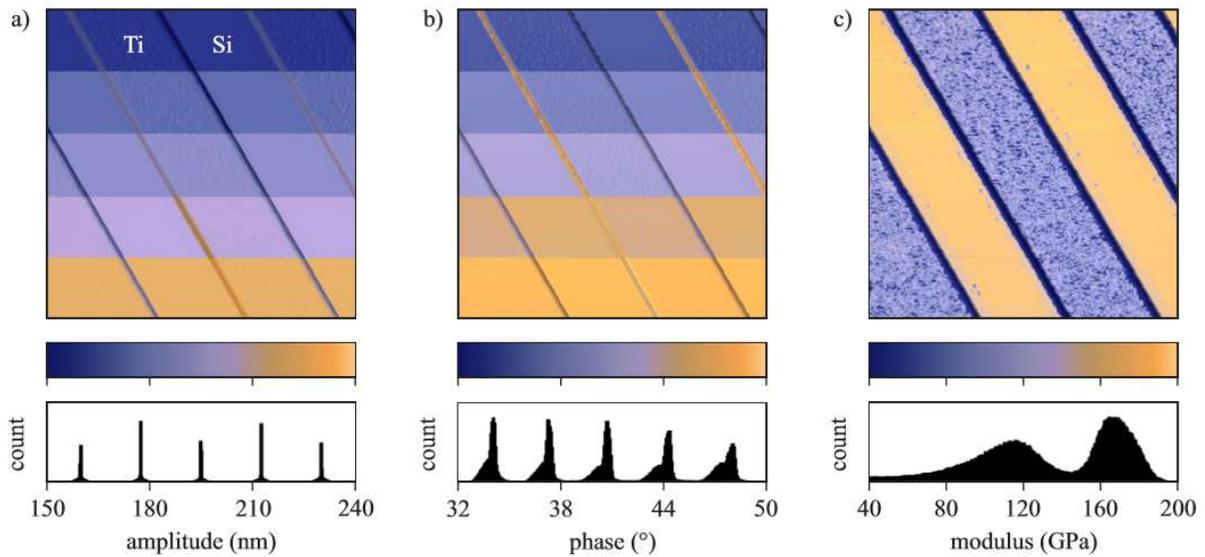

**Figure 4.** AM-FM results for a Si sample with thin-film stripes of Ti. (a) The amplitude setpoint channel for the first eigenmode was increased incrementally from 160 to 230 nm (free air amplitude $A_{1,\text{free}} = 320$ nm). (b) Phase image for the first eigenmode shows the resulting step increases corresponding to increases in the amplitude setpoint. (c) Elastic modulus image demonstrates the measured modulus is independent of amplitude setpoint. The indentation depths

were 1.7 nm on Si and 1.9 nm on Ti. Cantilever: Olympus AC160 with $k_1 \approx 26$ N/m, $f_1 \approx 300$ kHz, $k_2 \approx 475$ N/m and $f_2 \approx 1.7$ MHz. Scan size 10 μm.

**Molecular-Level Spatial Resolution**

In addition to quantitative measurements, AM-FM mode can be used for high resolution imaging. We demonstrate this ability by imaging the structure of semi-crystalline polymers on a molecular level. Although similar resolution has been reported previously,[34,35] it relied on custom cantilevers with specialized, extra-sharp tips operating in a torsional tapping mode. In contrast, for AM-FM imaging shown here, we used standard, commercially available cantilevers and obtained similar or better resolution images. Of course, AM-FM imaging also provides quantitative stiffness information alongside the topographic mapping.

It is important to note that for such high-resolution imaging, the continuum mechanics contact models used elsewhere in this work do not apply, mostly because the sample cannot be considered a semi-infinite half plane. Interpretation of the data in terms of modulus requires additional considerations beyond the scope of this article. Therefore, results in this section are presented in terms of the $\langle k_{ts} \rangle$. (Specifically, the reported tip-sample interaction stiffness in this section is the time-averaged stiffness change of the second eigenmode of the cantilever during intermittent contact experience while AM-FM imaging, as described in detail in Ref. 26)

In the first example, long (>500 nm) crystallites of syndiotactic polypropylene (sPP) formed during solvent evaporation while spin-coating were imaged in AM-FM mode. In the topography image shown in Figure 5a, the height difference between crystalline features is ~5 nm. Figure 5b shows a tip-sample interaction stiffness map of a single crystallite, where the crystal lattice of the polymer is clearly visible. The distance between the regular features parallel to the crystallite (indicated by the solid arrow in Figure 5b) was measured to be 1.14 ± 0.06 nm, while the distance between regular features perpendicular to the crystallite (dashed arrow in Figure 5b) was 0.76 ± 0.05 nm . (Measurement uncertainty represents one standard deviation in the individual measurements.) Additionally, FFT of the tip-sample interaction stiffness image confirms the periodicity of the observed features (see supplemental Materials figure S1). These values correspond closely to the x-ray crystallography values of b=1.12 nm and c= 0.74 nm predicted for the crystal lattice of sPP.[36]

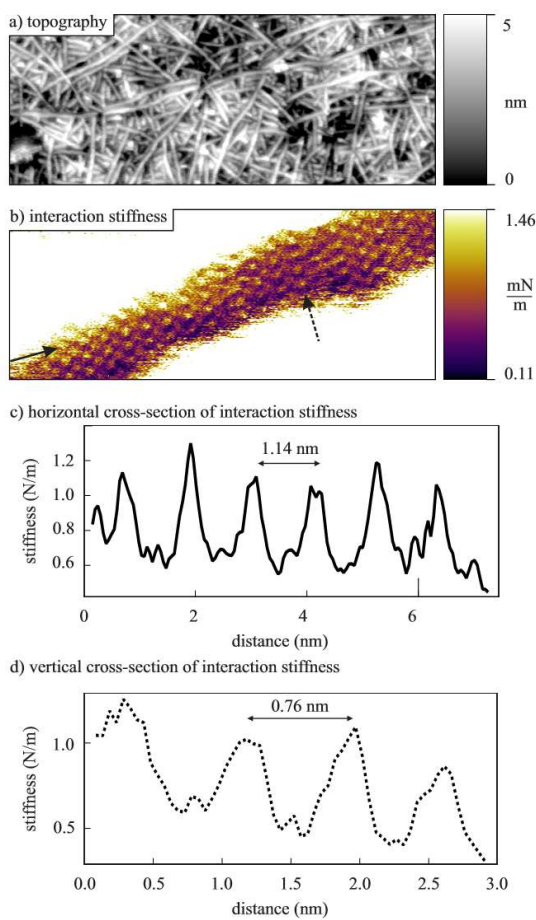

**Figure 5.** High spatial resolution imaging of syndiotactic polypropylene (sPP) in AM-FM mode. (a) Topography image of randomly-distributed sPP crystallites. (b) Stiffness image of a single

crystallite, where individual methyl groups of the polymer backbone are visible. (c) Line section along the crystallite in (b) in the direction of the solid arrow. The spacing between polymer chains is 1.14 ± 0.06 nm. (d) Line section along the polymer chain indicated by the dashed arrow in (b). The spacing between regular features is 0.76 ± 0.05 nm. Cantilever: NanoWorld ArrowUHF with $k_1 \approx 8.5$ N/m, $f_1 \approx 1.5$ MHz, $k_2 \approx 42$ N/m and $f_2 \approx 4.5$ MHz.

In the second example, high spatial resolution AM-FM imaging was demonstrated on polyethylene (PE). In Figure 6a, the topography image displays flat lamellae distinctive of crystalline PE. The tip-sample interaction stiffness image with smaller scan size in Figure 6b highlights two regions: (I) the interface between crystalline and amorphous phases of the polymer and (II) tight packing of polymer chains within the crystalline region.

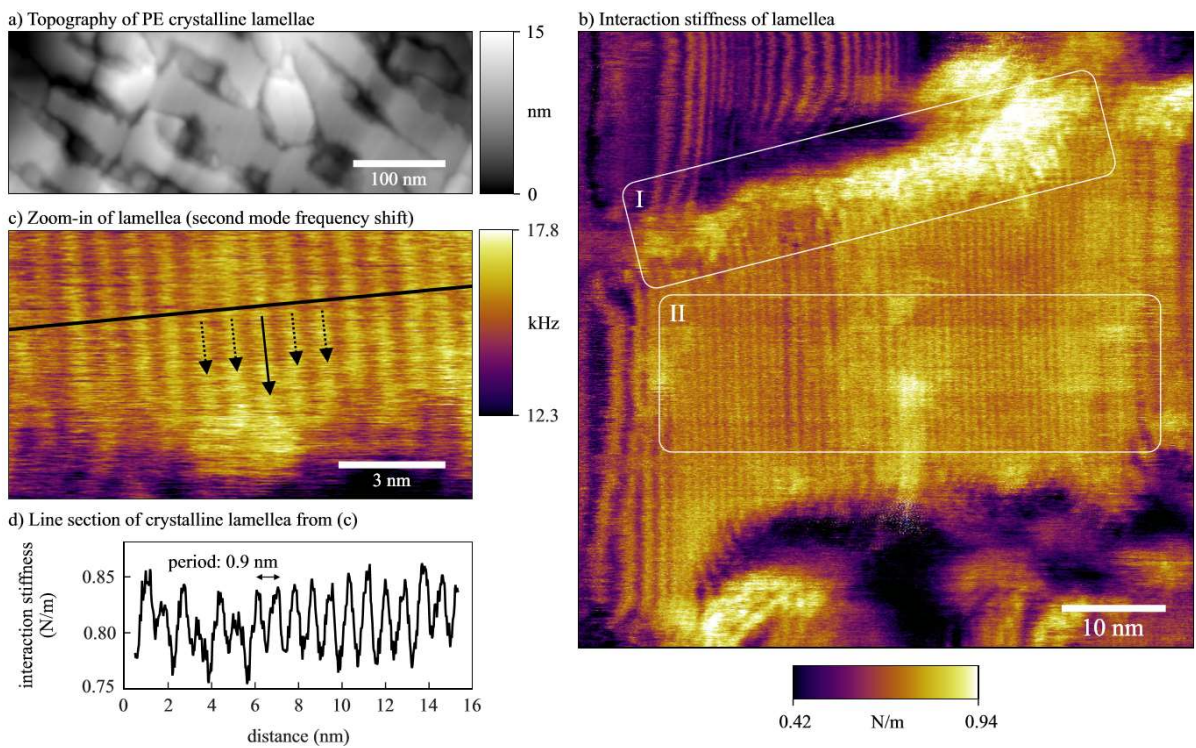

**Figure 6.** High spatial resolution imaging of polyethylene (PE). (a) Topography image of PE crystalline lamellae. (b) Tip-sample interaction stiffness image of a lamella with disordered end chains at the interface of the crystalline and amorphous phases (region I) and tightly packed polymer chains (region II). Blue pixels represent data points with NaN values, where negative indentation was calculated and was deemed unphysical. (c) Expanded image of tightly packed chains showing polymer chains looping out and back into the crystalline phase (solid arrows) and one of the polymer chains extending from crystalline into amorphous phase (dashed arrow). (d) Line section corresponding to the black line in (c) shows a 0.9 ± 0.1 nm spacing of polymer chains. See supplemental figure S3 for additional PE sample features such as terrace steps. Cantilever: NanoWorld ArrowUHF with $k_1 \approx 8.5$ N/m, $f_1 \approx 1.5$ MHz, $k_2 \approx 42$ N/m and $f_2 \approx 4.5$ MHz.

Figure 6c is a magnification of the interface region (region I in Figure 6b). Regularly-aligned, tightly-packed polymer chains are clearly visible. While most of the polymer chains loop out and back into the crystalline phase, some chains occasionally extend from the ordered crystalline phase into the disordered amorphous phase. This type of conformational arrangement of polymer chains has been predicted by theory and simulation[37] and is believed to be a factor in polymer crystallization that in turn affects bulk material properties. FFT of the tip-sample interaction stiffness image show a periodicity of 0.88 nm (see supplemental materials figure S2). Measurements of line sections such as the one in Figure 6d yield 0.9 ± 0.1 nm for the spacing between the tightly-packed chains, in good agreement with the value of 0.809 nm obtained by x-ray crystallography results for mechanically deformed PE samples.[38]

High-resolution AM-FM imaging can also be achieved in liquid environments, as demonstrated by the image of DNA in buffer solution in Figure 7. The double helix conformation of the DNA strands was resolved in the topographic and the interaction stiffness map of the AM-FM images. The 3.4 nm double helix pitch is clearly visible, with the major and minor groove widths of 2.2 nm and 1.2 nm, respectively, in the B conformation of the molecule. Ido et al.[39] have demonstrated very high resolution imaging of DNA topography in fluid, even beyond the helix resolution shown here. However, we have demonstrated quantitative characterization of the localized stiffness of the molecule, not only the topography. In addition, AM-FM imaging of biological materials has also been demonstrated in air.[40]

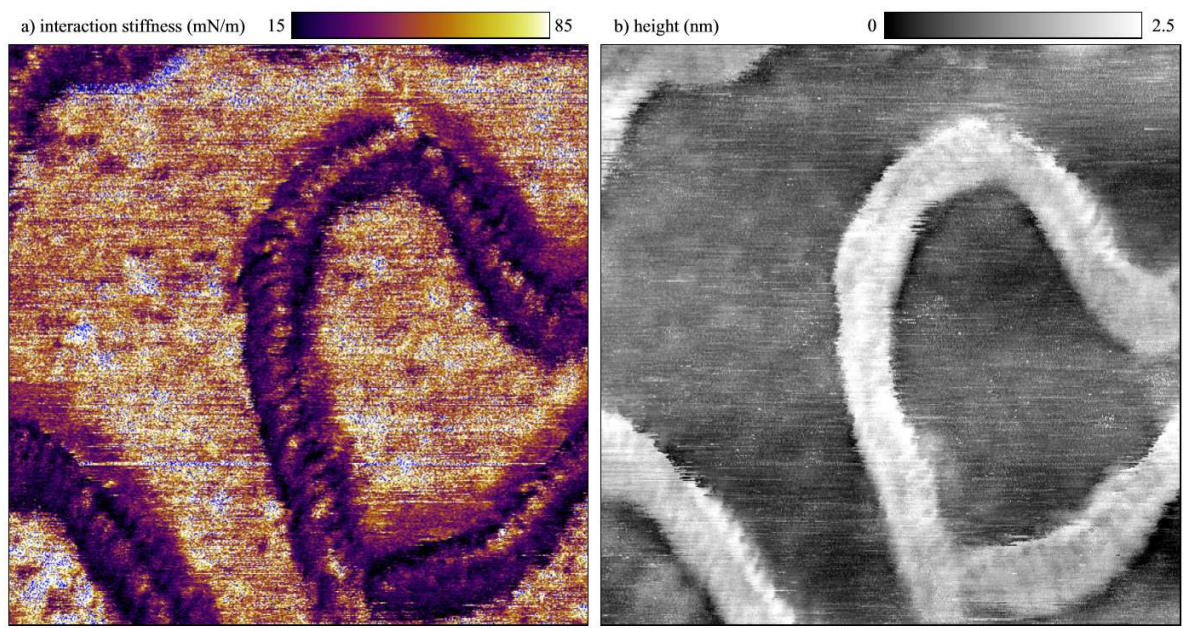

**Figure 7.** Molecular-resolution AM-FM interaction stiffness ($\langle k_{ts} \rangle$) map and topography of DNA immobilized on mica in $NiCl_2$ buffer. The double helix of the DNA strand is resolved. Here, $k_2$ was calibrated by assuming the stiffness-frequency power law relationship of an Euler-

Bernoulli beam.[28] Blue pixels represent data points with NaN values, where negative indentation was calculated and was deemed unphysical. Cantilever: Olympus AC40TS with $k_1 \approx 0.12$ N/m, $f_1 \approx 31$ kHz, $k_2 \approx 10.0$ N/m and $f_2 \approx 283$ kHz. Scan size 72 nm.

**Fast Nanomechanical Mapping**

Small cantilevers with resonant frequencies in the megahertz can be used to perform AM-FM imaging at very rapid scan rates while preserving nanomechanical information and spatial resolution. This imaging is much faster than other nanomechanical techniques such as force curve mapping, contact resonance AFM, and fast force curve mapping based on pulsed force AFM.[41,42]

Fast AM-FM imaging is demonstrated in Figure 8, which shows a PS/PP polymer film imaged at increasing line scan rates. As the line scan rate increased from 2 Hz to 39 Hz, the corresponding acquisition time for a complete 256×256 pixel image decreased from 128 s to 6.5 s. Despite the expected compromise between image quality and scanning speed, the modulus mapping remains relatively accurate even at very high speeds. Analysis of the histogram in Figure 8g indicates the measured modulus of PP is observed to increase by less than 10% for a 20× increase in scanning speed. We expect that imaging rates will increase as faster cantilevers and microscopes become available.

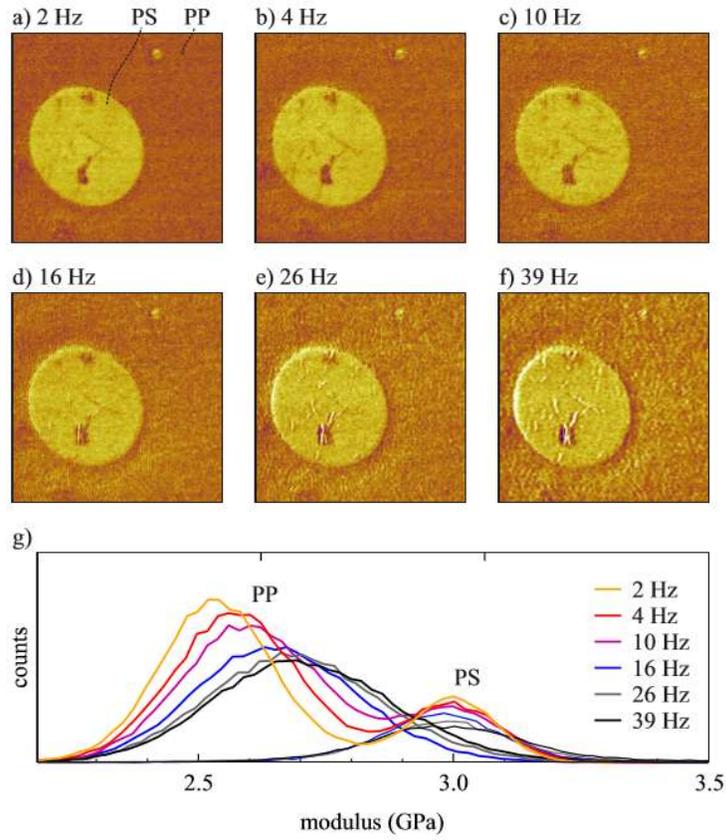

**Figure 8.** (a)-(f) Scan rate dependent AM-FM modulus images of a polystyrene/polypropylene (PS/PP) film obtained at increasing imaging speeds. The line scan rates of 2 Hz, 4 Hz, 10 Hz, 16 Hz, 26 Hz and 39 Hz correspond to acquisition times of 128 s, 66 s, 26 s, 16.5 s, 9.8 s and 6.5 s, respectively, for these 256×256 pixel images. (g) Histogram of all images demonstrates modulus changes with increasing imaging rates. A PS region of the sample was used as an internal reference with an assumed modulus of 3 GPa. This demonstrates the ability of the method to separate materials with very similar moduli at very rapid scan rates. Cantilever: NanoWorld ArrowUHF wiht $k_1 \approx 7$ N/m, $f_1 \approx 1.2$ MHz, $k_2 \approx 35$ N/m and $f_2 \approx 3.5$ MHz. Scan size 2.5 μm. The small change in the measured polypropylene modulus is directly attributable to small changes in the observables associated with increased feedback errors as the scan rate increases.

**SUMMARY**

Tapping mode is arguably the most successful and popular AFM imaging technique. Yet despite 30 years of use and development, quantitative, repeatable and accurate estimation of sample moduli from tapping mode imaging has not made it into the mainstream. This work demonstrates that the addition of extra observables from bimodal operation of the cantilever during imaging enables modulus mapping on a wide range of materials. Modulus mapping was performed with bimodal AM-FM techniques on samples ranging from compliant polymers with low modulus (~0.1-3 GPa) to very stiff metals with high modulus (>100 GPa). Notably, the results show that the same cantilever can quantitatively image modulus across three orders of magnitude, provided the cantilever tip size is calibrated on an appropriate reference sample. Furthermore, molecular-level spatial resolution was achieved with AM-FM imaging on polymer chains in ambient conditions and revealed chain spacing and conformation predicted by theory and other experimental methods. Lastly, the tapping-mode basis of AM-FM mode enables high imaging speeds: quantitative nanomechanical imaging was demonstrated at rates nearing 5 s per image with only slight degradation in modulus accuracy. With these capabilities for fast imaging, high-resolution nanomechanical mapping over a wide modulus range, AM-FM imaging is likely to prove a valuable nanoscale characterization tool for many applications.

**EXPERIMENTAL METHODS**

**Instrument**

All experiments were performed on Cypher S™ and Cypher-ES™ AFMs (Asylum Research – an Oxford Instruments Company, Santa Barbara, CA) equipped with blueDrive™ photothermal excitation. The cantilevers used for imaging were AC160 and AC40 (Olympus, Tokyo, Japan) and ArrowUHF (NanoWorld. Neuchâtel, Switzerland). All cantilevers had a reflective coating on the back side.

**Sample Preparation**

**a. Thin-film polymer blends**

Thin-film polymer blend samples of polystyrene/polypropylene (PS/PP) and polystyrene/polycaprolactone (PS/PCL) were prepared in-house by spin coating (Model WS-400BZ-6NPP/LITE, Laurell Technologies Corp., North Wales, PA) polymer solutions onto silicon wafers (orientation <100>, p-type/boron doped, Silicon Quest International, San Jose, CA). Polystyrene pellets (430102, Sigma-Aldrich, St. Louis, MO) and polypropylene pellets (452157, Sigma-Aldrich, St. Louis, MO) were dissolved in p-xylene (99%, Alfa Aesar, Ward Hill, MA) in 3:1 ratio. Polystyrene pellets and polycaprolactone pellets (440744, Sigma-Aldrich, St. Louis, MO) were dissolved in chloroform (Product #C2432, Sigma-Aldrich, St-Louis, MO) in 2:1 ratio. All the films were >30 nm in thickness.

**b. Polyethylene (PE)**

The PE sample used for high-resolution imaging was prepared by the procedure published by Mullin and Hobbs.[35] A small amount of 82.9 kDa polyethylene (PSS-pe83k, Polymer Standards Service USA, Amherst, MA) was heated on a glass slide and then mechanically sheared as the sample cooled

**c. Polydimethylsiloxane (PDMS)**

The PDMS sample was prepared by molding silicone elastomer (Sylgard 184 encapsulant, Dow Corning, Auburn, MI) at 1:10 ratio and curing at 60°C.

**d. Multilayer polymer packaging film**

The multilayer polymer sample was prepared by placing the film in a clamp inside a cryomicrotome (EM UC7 cryochamber, Leica Microsystems, Buffalo Grove, IL) at -120 °C. Slices ~100 nm thick were removed from the sample until a cross section of the material was obtained. The sample was then thawed under a flow of nitrogen and imaged. Experiments involving a reference sample used a 2×2 mm$^2$ piece of ultra-high-molecular-weight–polyethylene[43] (UHMWPE) that had been cryomicrotomed at -80 °C.

**e. Tin-lead (Sn/Pb) solder**

The solder sample was prepared by melting ~0.5 g of a 50:50 Sn/Pb solder wire (Alpha Fry, Cookson Electronics Assembly Materials, South Plainfield, NJ), depositing it on freshly cleaved mica sheet and heating the sample on a hot plate. Once the solder had melted, another mica sheet was used to press against the molten solder to obtain a flat sample. The mica was removed after the sample had cooled.

**f. Silicon-titanium (Si-Ti) grid**

A piece of Si wafer patterned with a ~200 nm thick Ti film[32] was cleaned by consecutive sonication in acetone (HPLC grade, Alfa Aesar, Ward Hill, MA), isopropanol (Gigabit, KMG

Chemicals Inc., Houston, TX) and ultrapure water (18.2 MΩ). The sample was dried with dry nitrogen and exposed to oxygen plasma for 1 min ( PE-IIA, Technics).

**g. DNA**

A stock solution of Lambda DNA BstE II Digest (D-9793,Sigma-Aldrich, St. Louis, MO) stored at 4°C in Tris-EDTA buffer at a concentration of 100 ng/µl was diluted to 10 ng/µl in a $NiCl_2$ buffer (10 mM $NiCl_2$, 40 mM Hepes, pH 6.8) immediately prior to use. 100 µl of this 10 ng/µl solution was applied on a freshly cleaved mica disc (diameter 1.5 cm). The mica disc had been previously attached to a metal puck using 24-hour epoxy (04005, Harman, Belleville, NJ). The solution of DNA was incubated for 10 min and thoroughly rinsed using a 2 mL syringe filled with de-ionized water and a needle. This pressured water flush insured attachment of DNA molecules on the surface. Finally, 150 µl of $NiCl_2$ buffer was added at the surface of the mica, and the preparation was then placed inside the AFM to be imaged.

**Conflict of interest**

The authors declare no competing financial interest.